\documentclass[letterpaper, 12 pt]{article}
\usepackage[a4paper, total={6in, 8in}]{geometry}
\usepackage{graphicx} 
\usepackage{amsmath}
\usepackage{graphicx}
\usepackage{cancel}
\usepackage{lipsum}
\usepackage{cite}
\usepackage{physics}
\usepackage{amsmath,amssymb}
\usepackage{dirtytalk}
\usepackage{xcolor}

\title{Acoustic Analogue for Quantum Field Theory with a Source term}
\author{ Akshat Pandey\footnote{apandey.physics@gmail.com} \\ \normalsize Department of Physics, Shiv Nadar Institution of Eminence \\ \normalsize Greater Noida, Uttar Pradesh-201314, India.}

\date{}
\begin{document}

\maketitle

\begin{abstract}
We propose a non-relativistic fluid analogue model for a scalar field coupled to a classical source. The generic analogue gravity model involves the phonon-field which is coupled to the acoustic metric. We work in the special relativity limit of the acoustic analogue. By assuming a time dependent external potential on the fluid system, we are able to model a source term for the scalar field. Upon quantisation, phonon creation due to the source is studied. 
\end{abstract}

\date{}

\maketitle

\section{Introduction}

Unruh, in 1981, predicted that in a non-relativistic fluid, a trans-sonic flow produces a thermal spectrum of phonons similar to Hawking radiation from black holes \cite{unruh}. Subsequently, several aspects of the analogy between fluid mechanics and systems coupled to gravity have been better understood and analogue gravity continues to be an active area of research \cite{review}. 

At the core of the analogy is the following principle \textbf{---} the equivalence between low-momentum phonon physics and quantum field theory in curved spacetime \cite{visser2}. As a consequence, extensive work has been done in constructing models analogous to the behaviour of quantum fields within black hole physics and cosmology \cite{ag1, ag2, ag3, ag4, ag5, ag6, ag7, ag8, ag9, ag10, ag11, ag12}. 

Some acoustic analogue models have also been constructed for quantum field theory (QFT) in flat spacetimes \cite{todd, other}. Todd \textit{et al.} explored acoustic analogues for Compton scattering \cite{todd}. They differentiated between in-universe observers namely ones coupled to the acoustic metric and who see acoustic Lorentz invariance, and other observes who uncoupled observers who do not. While in the present paper we will be concerned with the acoustic metric framework, there are other ways in which analogues models for QFT can be constructed using phonon physics. For example, Leizerovitch and Reznik \cite{other} have proposed an analogue model for Quantum Electrodynamics within Bose-Einstein condensates.

Within the acoustic metric analogy, choosing the background fluid to be stationary corresponds to an analogue Minkowski spacetime. Thus we end up with a massless, free Klein-Gordon (KG) theory. 

In this paper we wish to understand, in a top-down way, the limits of this fluid analogy to the KG field theory.
One can immediately see that the fluid analogy breaks down as soon as one tries to go beyond the free theory, as the phonon-interactions are not $\sim \phi^4$ or $\sim \phi^3+ \phi^4$. Thus including phonon interactions would lead to the violation of the analogue Lorentz symmetry \cite{vissersome}. Also, within this acoustic-metric framework, one cannot study interactions between two different fields as there is only one field coupled to the acoustic metric, namely that of sound.

However, there is one kind interaction of that both, has physical implications for the phonons in the fluid, and which can be included without disturbing the Lorentz invariance in the analogy. This is namely the interaction of the quantum field with an external classical source. In the rest of the paper we will see how assuming a time dependent external potential on the fluid system can be used to model a scalar field with a source term and how such a system leads to phonon creation in the vacuum upon quantisation.

This paper is organised as follows. In section 2, we revisit parts of Visser's derivation for the acoustic metric \cite{visser}. We modify a few physical assumptions to obtain a classical KG equation with a source term. In section 3, we explore some consequences of the system upon quantisation. We end with a summary in section 4.

\section{Massless scalar field equation with source}

We assume fluid to be barotropic, inviscid and irrotational. The dynamics is completely described by the continuity and Euler equations.

We start with the continuity equation given by
\begin{equation}
   \displaystyle\frac{\partial}{\partial t} \rho+\vec{\nabla} \cdot(\rho \vec{v})=0
   \label{1}
\end{equation}
The Euler equation of motion reads
\begin{equation}
    \rho \frac{\mathrm{d} v}{\mathrm{~d} t} \equiv \rho\left[\displaystyle\frac{\partial}{\partial t} \vec{v}+(\vec{v} \cdot \vec{\nabla}) \vec{v}\right]=\vec{F}
    \label{2}
\end{equation}
Since the fluid is inviscid the only terms of relevance in $\vec{F}$ are the ones of the pressure gradient and external potential $\chi$. For example, $\chi$ could be the gravitational potential. As such we can write
$
\vec{F}=-\nabla p-\rho \nabla \chi
$. Thus equation (\ref{2}) acquires the form
\begin{equation}
   \rho\left[\displaystyle\frac{\partial}{\partial t} \vec{v}+(\vec{v} \cdot \vec{\nabla}) \vec{v}\right]=-\nabla p-\rho \nabla \chi
   \label{3}
\end{equation}
Upon rearranging:
$$
\partial_{t} \vec{v}=-(\vec{v} \cdot \vec{\nabla}) \vec{v}-\frac{1}{\rho} \nabla p-\nabla\ \chi
$$
and using the identity for vector products \\
$$(\vec{v} \cdot \vec{\nabla}) \vec{v}=-\vec{v} \times(\nabla \times \vec{v})+\frac{1}{2}\nabla v^{2}$$ 
we are led to
\begin{equation}
\partial_{t} \vec{v}=\vec{v} \times(\nabla \times \vec{v})-\frac{1}{\rho} \nabla p-\nabla\left(\frac{1}{2} v^{2}+\chi\right)
\label{4}
\end{equation}\\

Since the fluid is irrorational which means $\nabla \times \vec{v}=0$,
we can define a velocity potential $ \phi$ in the manner $ \vec{v}= -\nabla\phi 
$ implying
\begin{equation}
-\partial_{t} \left(\nabla\phi\right) =-\frac{1}{\rho} \nabla p-\nabla\left(\frac{1}{2}  \left(\nabla\phi\right)^{2}+\chi\right)
\label{5}
\end{equation}\\
Employing the barotropic condition i.e. $ \rho \equiv \rho(p)$
enables us to define a quantity $h$ whose gradient has the integral representation 
\begin{equation}
b(p)=\int_{0}^{p} \frac{d p^{\prime}}{\rho\left(p^{\prime}\right)}
\label{6}
\end{equation}

Thus 
\begin{equation}
\nabla b=\frac{1}{\rho} \nabla p
\label{7}
\end{equation}

As a result the Euler equation becomes
\begin{equation}
    \nabla\left(-\partial_{t} \phi+b+\frac{1}{2}(\nabla \phi)^{2}+\chi \right)=0
    \label{8}
\end{equation}
which immediately suggests that
\begin{equation}
    -\partial_{t} \phi+b+\frac{1}{2}(\nabla \phi)^{2}+\chi = \Lambda
    \label{9}
\end{equation}
where the constant $\Lambda$ can be absorbed in $\chi$ resulting in a simpler version of the Euler equation

\begin{equation}
    -\partial_{t} \phi+b+\frac{1}{2}(\nabla \phi)^{2}+\chi = 0
    \label{10}
\end{equation}

Let us now adopt the well-known procedure of linearisation with respect to a small parameter $\epsilon$ in which the density $\rho$, pressure $p$, $\phi$ and $\chi$ are represented by

\begin{eqnarray}
 &&\rho=\rho_{0}+\epsilon \rho_{1}+O\left(\epsilon^{2}\right)
 \label{11}\\ 
 && p=p_{0}+\epsilon p_{1}+O\left(\epsilon^{2}\right)
 \label{12}\\
 && \phi= \phi_{0}+\epsilon \phi_{1}+O\left(\epsilon^{2}\right)
 \label{13}
 \end{eqnarray}

This leads to the linearisation of the barotropy relation 

The barotropic condition gives us
\begin{equation}
b(p)=h\left(p_{0}+\epsilon p_{1}+O\left(\epsilon^{2}\right)\right)=\int_{0}^{p+\epsilon p_{1}+O\left(\epsilon^{2}\right)} \frac{d p^{\prime}}{\rho\left(p^{\prime}\right)}=b(p_{0})+\epsilon \frac{p_{1}}{\rho_{0}+\epsilon \rho_{1}}+O\left(\epsilon^{2}\right)
\label{15}\\\\
\end{equation}
This can be written up to an order of $\epsilon$ 
\begin{equation}
 b_{0}+\frac{\epsilon p_{1}}{\rho_{0}+\epsilon \rho_{1}} \approx b_{0}+\frac{\epsilon p_{1}}{\rho_{0}}
 \label{16}\\
\end{equation}

Additionally, we also linearise the external potential 

 \begin{equation}
     \chi= \chi_{0}+\epsilon \chi_{1}+O\left(\epsilon^{2}\right)
 \label{14}
 \end{equation}

As we shall see, this perturbation in the external potential eventually leads to the presence of the source in the field equation.

Since we want to get to flat-spacetime QFT, at this point we can make some simplifying assumptions. We assume background fluid to be at rest i.e. $\vec{v}_0=0$ implying $\phi_0$ is a constant. Additionally we assume the background fluid density $\rho_0$ and pressure $p_0$ to be constant.

Following the linearisation, the continuity equation becomes

\begin{equation}
    \partial_t \rho_1 + \nabla . (\rho_0 \vec{v}_1) = 0
    \label{17}
\end{equation}

And the Euler equation takes the form

\begin{equation}
    -\partial_t \phi_1 + \frac{p_1}{\rho} + \chi_1 = 0
    \label{18}
\end{equation}

Since, fluid is barotropic, using equation (\ref{18}), $\rho_1$ takes the form

\begin{equation}
    \rho_1 = \frac{\partial \rho}{\partial p} \rho_0(\partial_t \phi_1 - \chi_1)
    \label{19}
\end{equation}

In such fluid systems, sound is defined as the a consequence of perturbation in the fluid quantities with the speed of sound given by

\begin{equation}
    c_s^{-2} \equiv \frac{\partial \rho}{\partial p}
    \label{20}
\end{equation}

Plugging this into the continuity equation, we get 

\begin{equation}
    \frac{1}{c_s^2} \partial_t(\partial_t \phi_1 - \chi_1) - \nabla^2 \phi_1 = 0
    \label{21}
\end{equation}

Upon rearranging terms, we get 

\begin{equation}
    \Box \phi_1 \equiv \left(\frac{1}{c_s^2} \partial_t^2 - \nabla^2\right) \phi_1 (\vec{x}, t) = \frac{\partial_t \chi_1 (\vec{x},t)}{c_s^2}
    \label{22}
\end{equation}

We see we have obtained the massless KG equation coupled to a classical source $j(\vec{x},t)$ defined as

\begin{equation}
    j(\vec{x},t) \equiv \frac{\partial_t \chi_1 (\vec{x},t)}{c_s^2}
    \label{23}
\end{equation}

We see that assuming that the external potential has a perturbation and that this perturbation is time-dependent leads to the source term. Further, one might think that the presence of only the time derivative in the source term could lead to inconsistencies with the analogue Lorentz symmetry. However, this choice works just fine as the system we are working with is completely non-relativistic and thus has a unique time coordinate \textbf{---} the laboratory time.

What we are interested in is the quantum version of equation (\ref{22}) which would then be the equation of motion for a phonon field \cite{visserbec, visser3}.  Note that since the source term is due to the external potential and not a part of the fluid system, it is still classical.

Thus we end up with a scalar quantum field coupled to a classical source. From here on, it is a standard textbook calculation \cite{coleman, peskin}. Nevertheless, in the next section we shall sketch out some results in particular those related to particle creation. We shall consider $\hbar = c_s = 1$. Further, to simplify notation we shall write $\phi_1$ as $\phi$.

\section{Particle creation}

The Lagrangian density corresponding to equation (\ref{22}) is

\begin{equation}
    \mathcal{L} = \frac{1}{2}\partial_\mu \phi \partial^\mu \phi + j\phi
    \label{24}
\end{equation}

As defined above, $j(x)$ is non-zero only for a finite time. We are interested in solutions after $j(x)$ has been turned off. The solution takes the following form

\begin{equation}
    \phi(x) = \phi_f(x) + \phi_s(x)
    \label{24}
\end{equation}

Here $\phi_f$ is the standard solution to the source free quantised KG equation that is

\begin{equation}
    \phi_f(x) = \int \frac{d^3 k}{(2 \pi)^3} \frac{1}{\sqrt{2 \omega_{\vec{k}}}}  
 (a_{\vec{k}}e^{ikx} + a_{\vec{k}}^{\dagger}e^{-ikx}) 
    \label{26}
\end{equation}

The addition to the solution can be written in term of the retarded Green function 
\begin{equation}
    \phi_s(x) = i \int d^4y D_R(x-y)j(y)
    \label{27}
\end{equation}

Since its all in the past, $D_R$ simplifies and we get

\begin{equation}
    \phi_s(x) = i \int d^4y \int \frac{d^3k}{(2 \pi)^3}\frac{1}{2 \omega_{\vec{k}}}(e^{-ik(x-y)}- e^{ik(x-y)})j(y)
    \label{28}
\end{equation}

Since the integral is over all $y$, we can make use of the Fourier transform of $j(y)$

\begin{equation}
    \Tilde{j}(k) = \int dy e^{iky} j(y)
    \label{29}
\end{equation}

Plugging this into equation (\ref{28}) we see equation (\ref{24}) takes the following form

\begin{equation}
    \phi(x) =  \int \frac{d^3 k}{(2 \pi)^3} \frac{1}{\sqrt{2 \omega_{\vec{k}}}}  
 \left(\left(a_{\vec{k}} + i\frac{\Tilde{j}(k)}{\sqrt{2 \omega_{\vec{k}}}}\right)e^{ikx} + \left( a_{\vec{k}}^{\dagger} - i\frac{ \Tilde{j}^*(k)}{\sqrt{2 \omega_{\vec{k}}}}\right) e^{-ikx}\right) 
    \label{30}
\end{equation}

The creation and annihilation operators have been shifted by a function. The new set of states constructed out of the shifted operators are the field theoretic versions of coherent states \cite{coherentreview}.

Therefore, we can define new set of creation and annihilation operators $A_{\vec{k}}$ and $A^{\dagger}_{\vec{k}}$ such that

\begin{equation}
    A_{\vec{k}} \equiv a_{\vec{k}} +  i\frac{\Tilde{j}(k)}{\sqrt{2 \omega_{\vec{k}}}}
    \label{31}
\end{equation}

In terms of these, the number operator $N$ can be written as

\begin{equation}
    N= \int \frac{d^3k}{(2 \pi)^3} A^{\dagger}_{\vec{k}}A_{\vec{k}}
    \label{32}
\end{equation}

For the source free theory, the number operator $N_f$ would simply be $A^{\dagger}_{\vec{k}}A_{\vec{k}} \xrightarrow{}a^{\dagger}_{\vec{k}}a_{\vec{k}}$ in the above equation. In that case by definition, the vacuum expectation value of $N_f$ is 0.

We are interested in calculating the vacuum expectation value of the number operator $N$

\begin{equation}
    \mel{0}{N}{0} = \int \frac{d^3k}{(2 \pi)^3} \frac{1}{2 \omega_{\vec{k}}}|\Tilde{j}(k)|^2
\end{equation}

Note that the vacuum state here is that of the free vacuum. We thus see the effects of the source on the free theory. We can check that upon considering the external potential to be time-independent $j$ and consequently $\Tilde{j}$ vanish. In this case equation (33) tells us that the particle number is 0. Thus, we see that it is the time dependent potential that leads to particle creation. These particles are indeed phonons given by the on-shell Fourier modes of the source.

We note that despite being a straightforward calculation, there are aspects of quantum fields with a classical source which are not well understood. Recently, Tinti \textit{et al.} highlighted the non-trivial dynamics of scattering processes in the presence of time dependent classical sources \cite{tinti}. Another interesting work in this context is by Fulling \textit{et al.} in which they explore the consistency between classical and quantum field theory using an external source \cite{fulling}.

Lastly, we want to emphasise that our analysis involved certain simplifying assumptions. To extend this to a physically testable situation these assumptions need to be carefully relaxed. Our model has finite size i.e. the integration over all $y$ is, in fact, over a finite region. This will lead to discretization in the $k$ values. More importantly, the $k$ values will have an associated UV cut-off, like in QFT. For the analogue model this becomes even more important because as mentioned in the introduction, the analogy with QFT in curved spacetimes holds only for low-momentum phonon physics.

\section{Summary}

In this paper, we presented a non-relativistic acoustic analogue model for a relativistic quantum field theory coupled to a classical source in flat spacetimes. We assumed the external potential acting on the fluid system to have a time dependent perturbation which, in turn, mimicked the classical source. We studied how this \say{source} in the analogue model would lead to the production of phonons in the fluid.

\end{document}